# Linear optical setups for active and passive quantum error correction in polarization encoded qubits

José Cláudio do Nascimento, Fábio Alencar Mendonça and Rubens Viana Ramos[•]

*Department of Teleinformatic Engineering, Federal University of Ceará, Campus do Pici, C.P. 6007, 60455-760, Fortaleza-Ceará, Brazil*

**Abstract**

In this work, we present active and passive linear optical setups for error correction in quantum communication systems that employ polarization of single-photon and mesoscopic coherent states. Applications in quantum communication systems are described.

The development of applications in quantum computation and quantum communication, like quantum teleportation [1] and quantum key distibution [2-4], has motivated a lot of efforts aiming to build reliable systems. The goal is to build systems able to correct the errors produced by the unavoidable quantum noise presence. In this direction, quantum error correction systems based on the introduction of redundancy on the information, through entangled states, have been proposed [5]. These codes have as objective to recover the information after its passage for the noisy channel; the errors are esteemed through the inserted redundancy, the same idea used in classical codes for classical communication. As

---

[•] Correspoding Author
*E-mail addresses*: rubens@deti.ufc.br, claudio@deti.ufc.br, alencar@deti.ufc.br

examples, one can quote the Shor's code [6] and the family of stabilized codes in the quantum limit of Hamming to error correction of the type bit-flip [7]. In contrast with these error correction systems based on entangled states, Kalamidas [8] proposed two simple optical schemes, one for quantum error rejection and the other for quantum error correction, using only linear optical devices. In stark contrast with others known proposals, his schemes do not require multi-photons entangled states and the error rejection and correction are not probabilistic. The main idea in Kalamidas' system is to realize, before lunching the polarization qubit through the channel, the transformation of the polarization qubit to horizontally polarized time-bin qubit. If the time separation of the time-bin qubit states is lower than the characteristic time of change of channel parameters, then both orthogonal states of the time-bin qubit will see the same channel and will suffer the same polarization transformation. This fact enables the error rejection/correction at the receiver, by realizing, in a smart way, the inverse transformation, from time-bin to polarization qubit.

Initially, let us discuss the single-photon linear-optical scheme for quantum error correction (quantum error rejection will not be discussed here) proposed in [8]. This scheme can be seen in Fig. 1.

The transmitter, Alice, has a single-photon in an unknown polarization state

$$|\Psi\rangle = \alpha|H\rangle + \beta|V\rangle, \qquad (1)$$

where $|H\rangle$ ($|V\rangle$) represents the horizontal (vertical) state. After the unbalanced polarization interferometer the state is

$$|\Psi\rangle = \alpha|H\rangle_S + \beta|V\rangle_L, \qquad (2)$$

since the horizontal component takes the short path, S, while the vertical component takes the long path, L. Alice turns on her Pockels cell only when *L*-path component is present, effecting the transformation $|V\rangle_L \to |H\rangle_L$. Hence, the state that Alice sends to Bob, the receiver, is

$$|\Phi\rangle = \alpha|H\rangle_S + \beta|H\rangle_L. \qquad (3)$$

Since the time separation between the components of the time-bin qubit is taken to be much lesser than the time of fluctuation of the channel parameters (fluctuation in the fiber birefringence) both components, S and L, will see the same stationary quantum channel modeled by the unitary operation *U*. The general form of *U* realizes the transformation

$$U|\Phi\rangle = \alpha\left(e^{i\lambda}\cos\varphi|H\rangle_S + e^{i\xi}\sin\varphi|V\rangle_S\right) + \beta\left(e^{i\lambda}\cos\varphi|H\rangle_L + \sin\varphi e^{i\xi}|V\rangle_L\right). \qquad (4)$$

where $\lambda$, $\xi$ and $\varphi$, random variables, are channel parameters. Equation (4) is the state that arrives at Bob's place. Bob, by its turn, turns on the Pockels cell $PC_B$ only when the S-path components of $U|\Phi\rangle$ are present, resulting the state

$$\alpha\left(e^{i\lambda}\cos\varphi|V\rangle_S + e^{i\xi}\sin\varphi|H\rangle_S\right) + \beta\left(e^{i\lambda}\cos\varphi|H\rangle_L + e^{i\xi}\sin\varphi|V\rangle_L\right). \qquad (5)$$

The state then propagates through a balanced polarization interferometer where $PC_{B(H)}$ is activate only when the S-path component is present and $PC_{B(V)}$, is activate only when the L-path component is present. The result is

$$\alpha\left(e^{i\lambda}\cos\varphi|V\rangle_S^1 + e^{i\xi}\sin\varphi|V\rangle_S^2\right) + \beta\left(e^{i\lambda}\cos\varphi|H\rangle_L^1 + e^{i\xi}\sin\varphi|H\rangle_L^2\right). \qquad (6)$$

At each output, 1 and 2, of Bob's balanced polarization interferometer, there is an unbalanced polarization interferometer, identical to Alice's one, followed by a HWP that rotates the polarization of π/2. After passing (6) by the unbalanced polarization interferometers, the state is

$$\alpha\left(e^{i\lambda}\cos\varphi|V\rangle_{SL}^{1}+e^{i\xi}sen\varphi|V\rangle_{SL}^{2}\right)+\beta\left(e^{i\lambda}\cos\varphi|H\rangle_{LS}^{1}+e^{i\xi}sen\varphi|H\rangle_{LS}^{2}\right). \quad (7)$$

Finaly, after the HWPs the Bob's final state is

$$|\Psi_{f}\rangle=e^{i\lambda}\cos\varphi\left(\alpha|H\rangle_{SL}^{1}+\beta|V\rangle_{LS}^{1}\right)+e^{i\xi}sen\varphi\left(\alpha|H\rangle_{SL}^{2}+\beta|V\rangle_{LS}^{2}\right) \quad (8)$$

Hence, the scheme of Fig. 1 enables each transmitted qubit to be obtained in an uncorrupted state. Each received qubit emerges randomly in either one of the two output modes (1 or 2) according to a distribution that depends on the channel parameter $\varphi$. A simplified version of that complex system that realizes the same error correction on the transmitted single-photon polarization encode qubit is shown in Fig. 2.

As can be seen in Fig. 2, the difference is Bob's setup. The state received by Bob is $U|\Phi\rangle$ as given by (4). The Pockels cell $PC_{B1}$ is activate only when the S-path component is present and $PC_{B2}$ is activate only when the L-path component is present. At each mode, 1 (upper arm) and 2 (lower arm) there exist an unbalanced polarization interferometer. In these interferometers, the horizontal component propagates through the long path while the vertical component propagates through the short path. When the corrupted state (4) arrives at Bob's place, after passing through the first PBS and Pockels cells, it is transformed to

$$\alpha\left(e^{i\lambda}\cos\varphi|H\rangle_S^2 + e^{i\xi}\sin\varphi|H\rangle_S^1\right) + \beta\left(e^{i\lambda}\cos\varphi|V\rangle_L^2 + e^{i\xi}\sin\varphi|V\rangle_L^1\right) \quad (9)$$

At last, passing through the unbalanced polarization interferometers, the final state is

$$|\Psi_f\rangle = e^{i\lambda}\cos\varphi\left(\alpha|H\rangle_{SL}^2 + \beta|V\rangle_{LS}^2\right) + e^{i\xi}\sin\varphi\left(\alpha|H\rangle_{SL}^1 + \beta|V\rangle_{LS}^1\right) \quad (10)$$

In (9) and (10) the superscripts 1 and 2 denote the paths in direction of the output modes 1 and 2. Since (10) and (8) are equal, both setup shown in Figs. 1 and 2 are equivalent, however, in the scheme here proposed one PBS, one $PC_B$ and two HWPs were taken away.

In both setups Bob obtains the uncorrupted state in mode 1 with probability $\cos^2(\varphi)$ and with probability $\sin^2(\varphi)$ in mode 2. When the channel is approximately an ideal channel Bob obtain the uncorrupted state more likely in mode 1, on the other hand, when $\varphi$ is allowed to vary over its total range of values according to a uniform distribution, then the probability of obtaining the uncorrupted state in either mode tends to 1/2. However, using an optical delay and an electro-optic switch to form a time multiplexing, one can have the state always at the same output (but in different times).

A polarimetric quantum key distribution setup can be directly implemented over a noisy channel of the type described by (4) just employing the error correction scheme shown in Fig. 2. In this case, Bob has firstly to inform to Alice in which time slots he received the corrected polarization (detection in time *SL*) and, for this smaller set, to inform to Alice the bases used by him. Hence, for the noisy channel having the parameter $\varphi$ uniformly distributed, not considering classical error correction and privacy amplification, in average

25% of the bits sent by Alice will be useful to form the key. The use of the error correction scheme not only permits a higher useful bit transmission rate but it has also an impact on the security. As is largely known, the errors produced by Eva are masked by the errors due to imperfection of the QKD setup. Hence, if these errors are minored, Eva becomes more visible. In order to analyze the security we will consider the Fuchs-Peres-Brandt individual attack [9,10]. Basically, if true single-photons are used (hence photon number attack is not possible), the most general attack that can be realized by Eve consists in to entangle the photon sent by Alice with another photon, provided by her, through a unitary operation. Eve then sends Alice's photon to Bob, and performs a measurement on the photon that she kept. In order to maximize the amount of Rényi information about the error-free sifted bits that Bob receives for a given level of disturbance (probability to cause an error in a sifted bit), Eve has to choose properly the quantum state of her photon, the unitary transformation and the POVM she will use. It has been proved the choices presented in Fig. 3 and in (11)-(15) are the optimal ones [9,10].

$$|e\rangle = C|+\rangle + S|-\rangle \tag{11}$$

$$C = \sqrt{1 - 2P_E}, \quad S = \sqrt{2P_E} \tag{12}$$

$$|\pm\rangle = (|0\rangle \pm |1\rangle)/\sqrt{2} \tag{13}$$

$$|0\rangle = \cos(\pi/8)|H\rangle + \sin(\pi/8)|V\rangle \tag{14}$$

$$|1\rangle = -\sin(\pi/8)|H\rangle + \cos(\pi/8)|V\rangle \tag{15}$$

In (12) $P_E$ is the probability of Eve's action to cause an error in Bob. The basis $\{|0\rangle,|1\rangle\}$ in (14)-(15) is the measurement basis used by Eve. The states $|\pm\rangle_R$ are given by $(|H\rangle \pm |V\rangle)/2^{1/2}$. For each possible choice of Alice, the Alice-Eve joint state is

$$U|H\rangle|e\rangle = |H\rangle|T_-\rangle + |V\rangle|T_E\rangle \tag{16}$$
$$U|V\rangle|e\rangle = |V\rangle|T_+\rangle + |H\rangle|T_E\rangle \tag{17}$$
$$U|+\rangle_R|e\rangle = |+\rangle_R|T_+\rangle + |-\rangle_R|T_E\rangle \tag{18}$$
$$U|-\rangle_R|e\rangle = |-\rangle_R|T_-\rangle + |+\rangle_R|T_E\rangle \tag{19}$$

where $|T_\pm\rangle$ and $|T_E\rangle$ are the (unnormalized) states

$$|T_\pm\rangle = C|+\rangle \pm (S/\sqrt{2})|-\rangle \tag{20}$$
$$|T_E\rangle = (S/\sqrt{2})|-\rangle. \tag{21}$$

As can be seen from (16)-(19), the probability of an error taking place, for $X \in \{H,V,+,-\}$ is equal to $\langle X^\perp, T_E|U|X,e\rangle = \langle X^\perp, T_E|X,T_\pm\rangle + \langle X^\perp, T_E|X^\perp,T_E\rangle = \langle T_E|T_E\rangle = S^2/2 = P_E$. It is easy to check that the probability of Eve getting the right result in her measurement, for each photon sent by Alice, is $0.5(1+2^{1/2}CS)$. If Eve chooses $P_E$ equal to 0 or 0.5, then the probability of Eve gets the right bit value is 0.5. For $P_E=0.25$, the maximal success probability is achieved, 0.8535.

For the security analysis, one must consider that Eve has the same encoder and decoder used by Alice and Bob, respectively. Eve decodes the encoded quantum state sent by Alice recovering with very high probability (since Eve is located very close to Alice) the correct polarization state. Then, she performs her attack, encodes the photon and sends it to Bob. If the state sent by Alice was, without loss of generality, $|+\rangle_R$, then we have the following sequence:

1. State after Eve's encoder (or at optical fiber input)

$$\frac{(|H\rangle_S + |H\rangle_L)}{\sqrt{2}}|T_+\rangle + \frac{(|H\rangle_S - |H\rangle_L)}{\sqrt{2}}|T_E\rangle \tag{22}$$

2. State after optical fiber propagation (or at Bob's input)

$$\left\{\frac{\left[\cos(\varphi)e^{i\lambda}|H\rangle_S+\sin(\varphi)e^{i\xi}|V\rangle_S\right]}{\sqrt{2}}+\frac{\left[\cos(\varphi)e^{i\lambda}|H\rangle_L+\sin(\varphi)e^{i\xi}|V\rangle_L\right]}{\sqrt{2}}\right\}|T_+\rangle+$$

$$\left\{\frac{\left[\cos(\varphi)e^{i\lambda}|H\rangle_S+\sin(\varphi)e^{i\xi}|V\rangle_S\right]}{\sqrt{2}}-\frac{\left[\cos(\varphi)e^{i\lambda}|H\rangle_L+\sin(\varphi)e^{i\xi}|V\rangle_L\right]}{\sqrt{2}}\right\}|T_E\rangle \quad (23)$$

3. State after the first PBS at Bob's place

$$\frac{\left[\cos(\varphi)e^{i\lambda}|H\rangle_S^2+\cos(\varphi)e^{i\lambda}|H\rangle_L^2\right]}{\sqrt{2}}|T_+\rangle+\frac{\left[\cos(\varphi)e^{i\lambda}|H\rangle_S^2-\cos(\varphi)e^{i\lambda}|H\rangle_L^2\right]}{\sqrt{2}}|T_E\rangle+$$

$$\frac{\left[\sin(\varphi)e^{i\xi}|V\rangle_S^1+\sin(\varphi)e^{i\xi}|V\rangle_L^1\right]}{\sqrt{2}}|T_+\rangle+\frac{\left[\sin(\varphi)e^{i\xi}|V\rangle_S^1-\sin(\varphi)e^{i\xi}|V\rangle_L^1\right]}{\sqrt{2}}|T_E\rangle \quad (24)$$

4. State after Pockels cells $PC_{B1}$ and $PC_{B2}$

$$\frac{\left[\cos(\varphi)e^{i\lambda}|H\rangle_S^2+\cos(\varphi)e^{i\lambda}|V\rangle_L^2\right]}{\sqrt{2}}|T_+\rangle+\frac{\left[\cos(\varphi)e^{i\lambda}|H\rangle_S^2-\cos(\varphi)e^{i\lambda}|V\rangle_L^2\right]}{\sqrt{2}}|T_E\rangle+$$

$$\frac{\left[\sin(\varphi)e^{i\xi}|H\rangle_S^1+\sin(\varphi)e^{i\xi}|V\rangle_L^1\right]}{\sqrt{2}}|T_+\rangle+\frac{\left[\sin(\varphi)e^{i\xi}|H\rangle_S^1-\sin(\varphi)e^{i\xi}|V\rangle_L^1\right]}{\sqrt{2}}|T_E\rangle \quad (25)$$

5. State after Bob's polarization interferometer (or at Bob's output)

$$|\Psi\rangle=e^{i\lambda}\cos(\varphi)\left\{\frac{\left[|H\rangle_{SL}^2+|V\rangle_{LS}^2\right]}{\sqrt{2}}|T_+\rangle+\frac{\left[|H\rangle_{SL}^2-|V\rangle_{LS}^2\right]}{\sqrt{2}}|T_E\rangle\right\}+$$

$$e^{i\xi}\sin(\varphi)\frac{\left[|H\rangle_{SL}^1+|V\rangle_{LS}^1\right]}{\sqrt{2}}|T_+\rangle+\frac{\left[|H\rangle_{SL}^1-|V\rangle_{LS}^1\right]}{\sqrt{2}}|T_E\rangle \quad (26)$$

$$|\Psi\rangle=e^{i\lambda}\cos(\varphi)\left[|+\rangle_R^2|T_+\rangle+|-\rangle_R^2|T_E\rangle\right]+e^{i\xi}\sin(\varphi)\left[|+\rangle_R^1|T_+\rangle+|-\rangle_R^1|T_E\rangle\right] \quad (27)$$

Observing (27) one can conclude that, in those cases where should not have any error (detection in time *SL*) there will be an error probability equal to $P_E$ exclusively due to Eve's action. The same will happen for all other possible choices of Alice. Thus, the error correction setup has no influence on the errors caused by Eva and, hence, Eve's action is more visible.

The main problem with setups shown in Figs. 1 and 2 is the synchronization of the Pockels cells. A passive version (a setup without using Pockels cells) of the error reject optical setup proposed in [8] was presented in [11]. The passive version can be used in optical systems employing single-photon and coherent states, however, in the former, the performance is not good because of the losses introduced by fiber couplers. Following the idea used in [11], we propose a passive optical setup for error correction. This is important because some quantum communication schemes using polarization of mesoscopic coherent states have been proposed [12, 13]. The passive, in the sense that none external action is needed, error correction setup for coherent state-based optical systems is shown in Fig. 4.

In Fig. 4 BS is a fiber coupler having transmittance $T = 1/\sqrt{2}$ and reflectance $R = i/\sqrt{2}$. The input state is the two-mode coherent state $|\alpha,\beta\rangle$, where the former (last) is the horizontal (vertical) part. The components of the input state are separated by the first PBS. The horizontal component takes the short path, receiving a phase shift of $-\pi/2$, while the vertical component takes the long path, passing trough the HWP, resulting in the state $|-i\alpha,0\rangle_S \otimes |\beta,0\rangle_L$. After the passage for BS, both pulses are launched in the quantum channel, separated by a time interval short enough in order to make the channel the same for both of them. The state at the channel input is $|\alpha/\sqrt{2},0\rangle_S \otimes |\beta/\sqrt{2},0\rangle_L$. As before, after channel propagation the state is:

$$\left|\frac{\alpha}{\sqrt{2}}\cos(\varphi)e^{i\lambda}, \frac{\alpha}{\sqrt{2}}\sin(\varphi)e^{i\xi}\right\rangle_S \otimes \left|\frac{\beta}{\sqrt{2}}\cos(\varphi)e^{i\lambda}, \frac{\beta}{\sqrt{2}}\sin(\varphi)e^{i\xi}\right\rangle_L. \qquad (28)$$

where once more $\varphi$, $\lambda$ and $\xi$ are the parameters of a general unitary transformation. When the *S* and *L* pulses arrive at the receiver, its components are separated by the PBS, resulting in the following state:

$$\left|\frac{\alpha}{\sqrt{2}}\cos(\varphi)e^{i\lambda},0\right\rangle^2_S \otimes \left|0,\frac{\alpha}{\sqrt{2}}\sin(\varphi)e^{i\xi}\right\rangle^1_S \otimes \left|\frac{\beta}{\sqrt{2}}\cos(\varphi)e^{i\lambda},0\right\rangle^2_L \otimes \left|0,\frac{\beta}{\sqrt{2}}\sin(\varphi)e^{i\xi}\right\rangle^1_L. \quad (29)$$

where the vertical components takes path 1 and horizontal component take path 2. In path 1 (2) there is a polarization rotator with angle $-\pi/4\,(\pi/4)$. After passing by the polarization rotators, the state is

$$\left|\frac{\alpha}{2}\cos(\varphi)e^{i\lambda},\frac{\alpha}{2}\cos(\varphi)e^{i\lambda}\right\rangle^2_S \otimes \left|\frac{\beta}{2}\cos(\varphi)e^{i\lambda},\frac{\beta}{2}\cos(\varphi)e^{i\lambda}\right\rangle^2_L \otimes$$
$$\left|\frac{\alpha}{2}\sin(\varphi)e^{i\xi},\frac{\alpha}{2}\sin(\varphi)e^{i\xi}\right\rangle^1_S \otimes \left|\frac{\beta}{2}\sin(\varphi)e^{i\xi},\frac{\beta}{2}\sin(\varphi)e^{i\xi}\right\rangle^1_L. \quad (30)$$

At the unbalanced polarization interferometers at Bob's side, the vertical component takes the short path while the horizontal component takes the long path. Hence, the final states at outputs 1 and 2 are

$$\left|\Psi^1_f\right\rangle = \left|0,\frac{\alpha}{2}\sin(\varphi)e^{i\xi}\right\rangle^1_{SS} \otimes \left|\frac{\alpha}{2}\sin(\varphi)e^{i\xi},\frac{\beta}{2}\sin(\varphi)e^{i\xi}\right\rangle^2_{SL} \otimes \left|\frac{\beta}{2}\sin(\varphi)e^{i\xi},0\right\rangle^2_{LL} \quad (31)$$

$$\left|\Psi^2_f\right\rangle = \left|0,\frac{\alpha}{2}\cos(\varphi)e^{i\lambda}\right\rangle^2_{SS} \otimes \left|\frac{\alpha}{2}\cos(\varphi)e^{i\lambda},\frac{\beta}{2}\cos(\varphi)e^{i\lambda}\right\rangle^2_{SL} \otimes \left|\frac{\beta}{2}\cos(\varphi)e^{i\lambda},0\right\rangle^2_{LL}. \quad (32)$$

As can be seen in (31)-(32), the second pulse (*SL*) has the correct polarization in both outputs. As stated before, the advantage of this scheme is its passive operation; on the other hand, its disadvantage is the loss of optical power in not useful pulses. For a lossless channel, the optical power of the useful pulse is $\sin^2(\varphi)/4$ in the path 1 and $\cos^2(\varphi)/4$ in the path 2 of the input pulse power at Alice's place.

In order to show the usefulness of the error correction scheme of Fig. 4, let us consider the setup presented in Fig. 5. It is a different version of the setup proposed in [12,13] including the error correction scheme of Fig. 4. For the setup of Fig. 5 only linear polarization is used. The state generated by Alice is given by

$$R(\theta_A)|\alpha,0\rangle = \exp\left(-\theta_A\left(\hat{a}_V^+\hat{a}_H - \hat{a}_H^+\hat{a}_V\right)\right)|\alpha,0\rangle = |\alpha\cos(\theta_A), \alpha\sin(\theta_A)\rangle, \quad (33)$$

where $\hat{a}_V$ and $\hat{a}_H$ are, respectively, the annihilation operators of the vertical and horizontal modes. For this linearly polarised light, the mean values and the variances of the Stokes parameters are

$$\langle S_1\rangle = |\alpha|^2\cos(2\theta); \langle S_2\rangle = |\alpha|^2\sin(2\theta); \langle S_3\rangle = 0 \quad (34)$$
$$V_{S_1} = V_{S_2} = V_{S_3} = |\alpha|^2 \quad (35)$$

An important question is how good one can distinguish between two linear polarisation states having a dephasing of $\theta$ between them. This measure is given by the inner product and, without loss of generality, let us consider one of the polarisations to be the linear horizontal state. The inner product is then given by:

$$D = |\langle\alpha,0|R(\theta)|\alpha,0\rangle|^2 = \exp\left(-2|\alpha|^2\sin^2(\theta)\right). \quad (36)$$

Hence, the smaller the angle $\theta$ the larger the necessary mean photon number of the coherent state for a good distinguishability. This point is crucial for the security of the quantum communication protocols based on mesoscopic coherent states, since it states that some relation between the mean photon number and the number of basis used must be obeyed in order to guarantee the security of the protocol. At this work we are not concerned with security issues, our aim is to show that the error correction scheme of Fig. 4 is useful in such quantum communication schemes. The quantum communication protocol works as follows: Alice chooses two polarization angles: $\phi_a^{bit} \in \{0, \pi/2\}$ for the bit, and $\phi_a^{basis} \in \{1,...,M\}$ where $M$ is the (odd) number of bases. Hence, the quantum state of the pulse sent by Alice is that one in (33) having $\theta_A = \phi_a^{bit} + \phi_a^{basis}$. After propagation in the quantum error correction scheme/optical channel, the state arriving at Bob's polarization rotator at time $LS+\tau$ (due to the optical delay) and $LS$ (the pulses at times $SS$ and $LL$ are discarded) are, respectively:

$$\left|\Psi_f^1\right\rangle = \left|\frac{\alpha\cos(\theta_A)}{2}\sin(\varphi)e^{i\xi}, \frac{\alpha\sin(\theta_A)}{2}\sin(\varphi)e^{i\xi}\right\rangle_{SL}^{1} \tag{37}$$

$$\left|\Psi_f^2\right\rangle = \left|\frac{\alpha\cos(\theta_A)}{2}\cos(\varphi)e^{i\lambda}, \frac{\alpha\sin(\theta_A)}{2}\cos(\varphi)e^{i\lambda}\right\rangle_{SL}^{2} \tag{38}$$

After Bob's action, the states are

$$\left|\Psi_f^1\right\rangle = \left|\frac{\sin(\varphi)e^{i\xi}}{2}\alpha\cos(\theta_A+\theta_B), \frac{\sin(\varphi)e^{i\xi}}{2}\alpha\sin(\theta_A+\theta_B)\right\rangle_{SL}^{1} \tag{39}$$

$$\left|\Psi_f^2\right\rangle = \left|\frac{\cos(\varphi)e^{i\lambda}}{2}\alpha\cos(\theta_A+\theta_B), \frac{\cos(\varphi)e^{i\lambda}}{2}\alpha\sin(\theta_A+\theta_B)\right\rangle_{SL}^{2} \tag{40}$$

The values chosen by Bob for $\theta_B$ are those previously agreed to Alice, $\theta_B = -\phi_A^{basis}$, as is required by the protocol [12,13]. Hence, the states arriving at Bob's measurer (last PBS + PIN-based photodetectors) are

$$\left|\Psi_f^1\right\rangle = \left|\frac{\sin(\varphi)e^{i\xi}}{2}\alpha\cos(\phi_A^{bit}), \frac{\sin(\varphi)e^{i\xi}}{2}\alpha\sin(\phi_A^{bit})\right\rangle_{SL}^1 = \begin{cases} \left|0.5\sin(\varphi)e^{i\xi}\alpha, 0\right\rangle & \text{if } \phi_A^{bit} = 0 \\ \left|0, 0.5\sin(\varphi)e^{i\xi}\alpha\right\rangle & \text{if } \phi_A^{bit} = \frac{\pi}{2} \end{cases} \quad (41)$$

$$\left|\Psi_f^2\right\rangle = \left|\frac{\cos(\varphi)e^{i\lambda}}{2}\alpha\cos(\phi_A^{bit}), \frac{\cos(\varphi)e^{i\lambda}}{2}\alpha\sin(\phi_A^{bit})\right\rangle_{SL}^2 = \begin{cases} \left|0.5\cos(\varphi)e^{i\lambda}\alpha, 0\right\rangle & \text{if } \phi_A^{bit} = 0 \\ \left|0, 0.5\cos(\varphi)e^{i\lambda}\alpha\right\rangle & \text{if } \phi_A^{bit} = \frac{\pi}{2} \end{cases} \quad (42)$$

Hence, the bit sent by Alice is the same that Bob measures. As can be seen from (37)-(42), the error correction scheme makes Bob to receive two copies of the state sent by Alice. Measuring their power, Bob can estimate the value of $\varphi$.

In summary, we began explaining the single-photon error correction setup using linear optics proposed in [8]. Following, we proposed a simplified version, in the sense that a lower number of optical devices is necessary, able to realize the same error correction. Its functioning was described. After that, we showed the security analysis of a polarimetric quantum key distribution system, using the proposed error correction scheme, when the eavesdropper attacks using the Fuch-Peres-Brandt attack. In this case, the error correction scheme corrects channel's errors but has no influence in the error due to Eve. Hence, Eve's action becomes more visible. Following, we proposed a passive error correction setup for coherent state-based polarimetric quantum communication. The setup proposed has the advantage of needing no control and, hence, it has an easy implementation. Its disadvantage is the loss of optical power in not useful pulses generated at the beam splitter. At last, it was explained how to use the proposed error correction system in a quantum communication setup employing mesoscopic coherent states.

**Acknowledgements**

This work was supported by the Brazilian agency FUNCAP.

# FIGURE 1

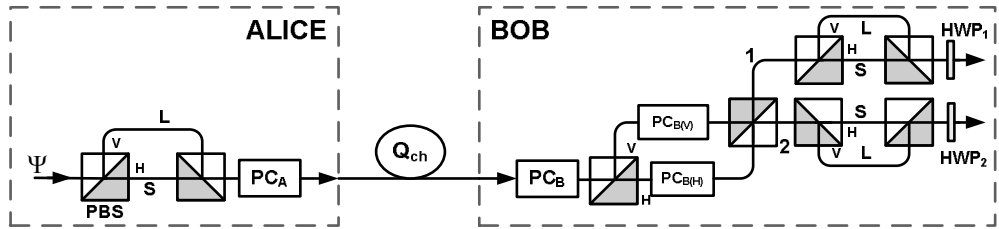

Figure 1 - Optical scheme for single-photon quantum error correction. PBS (Polarization beam splitter), PC (Pockels cell), HWP (Half-wave plate).

**FIGURE 2**

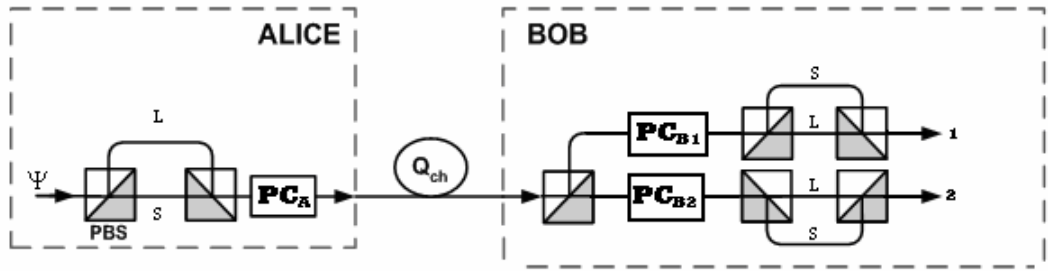

Figure 2 - Simplified scheme for single-photon quantum error correction.

**FIGURE 3**

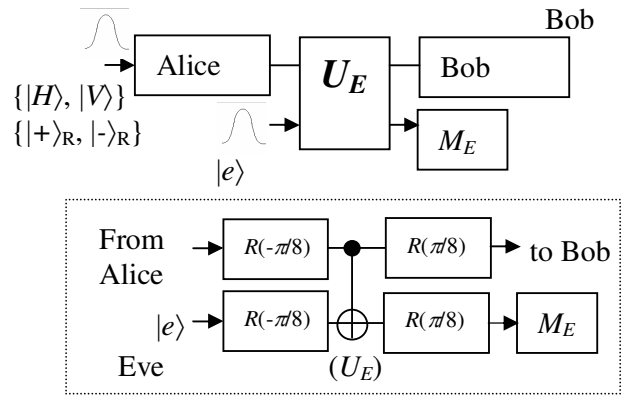

Figure 3 - Eve using Fuchs-Peres-Brandt attack. $M_E$ is a measurer and $R$ are polarization rotators.

# FIGURE 4

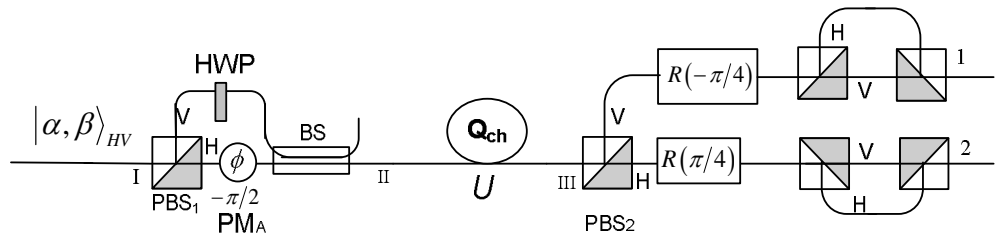

Figure 4 - Passive error correction setup for coherent state-based quantum communication. $R$ is a polarization rotator, PM is a phase modulator and BS is a 50/50 fiber coupler.

**FIGURE 5**

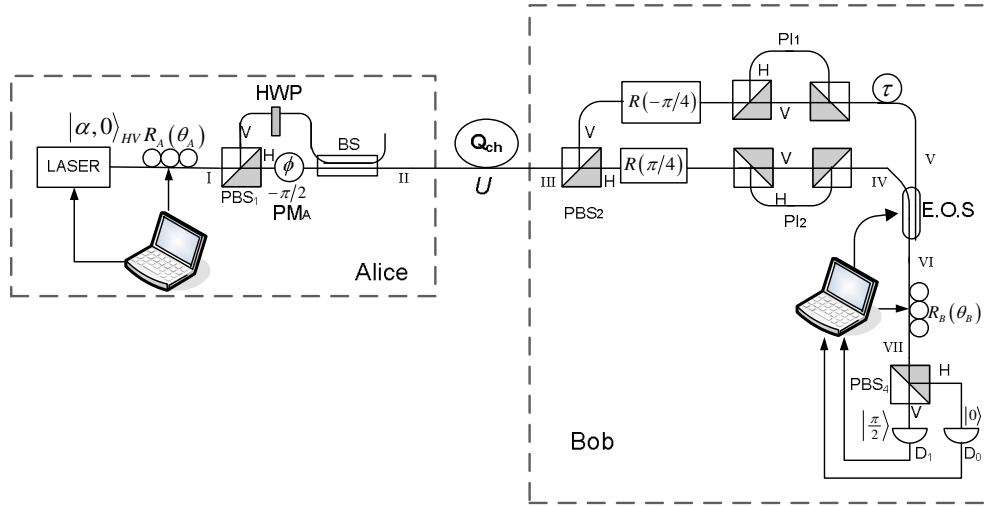

Figure 5: Optical setup for quantum communication using mesoscopic coherent state employing passive quantum error correction. E.O.S – Electric-optical switch, $PM_A$ – Alice's Phase modulator, $D_{0,1}$ – PIN-based photodetectors.